\newcommand{\version}{March 8, 2001}

\documentclass[12pt]{article}

\usepackage{a4,amsthm,amsfonts,latexsym,amssymb}

{\catcode `\@=11 \global\let\AddToReset=\@addtoreset}
\AddToReset{equation}{section}

\swapnumbers
\theoremstyle{plain}
\newtheorem{thm}{THEOREM}[section]
\newtheorem{cl}[thm]{COROLLARY}
\newtheorem{lem}[thm]{LEMMA}
\newtheorem{prop}[thm]{PROPOSITION}
\theoremstyle{definition}
\newtheorem{rem}[thm]{Remark}

\newcommand{\beq}{\begin{equation}}
\newcommand{\eeq}{\end{equation}}
\def\beqa{\begin{eqnarray}}
\def\eeqa{\end{eqnarray}}

\newcommand{\R}{{\mathbb R}}
\newcommand{\Rm}{{\mathcal R}}
\newcommand{\C}{{\mathbb C}}
\newcommand{\N}{{\mathbb N}}
\newcommand{\D}{{\mathcal D}}
\newcommand{\Ll}{{\mathcal L}}

\newcommand{\hp}{ h_{\xpp}}

\newcommand{\eps}{\varepsilon}
\newcommand{\A}{{\bf A}}

\newcommand{\x}{{\bf x}}
\newcommand{\y}{{\bf y}}

\newcommand{\xperp}{\x_\perp}

\newcommand{\Tr}{{\rm Tr}}
\newcommand{\T}{{\rm Tr}_{{\mathcal {L}}^2 ({\mathbb {R}})}}

\newcommand{\rs}{\rho^{\rm STF}}
\newcommand{\ri}{\rho^{(i)}}
\newcommand{\rd}{\rho^{\rm DSTF}}
\newcommand{\rmd}{\rho^{\rm MSTF}}

\newcommand{\Ed}{{E}^{\rm DSTF}}

\newcommand{\E}{{\mathcal E}^{\rm DSTF}}
\newcommand{\Edo}{{\mathcal E}^{\rm 1DSTF}}
\newcommand{\Ema}{{\mathcal E}}
\newcommand{\Em}{{\mathcal E}^{\rm MSTF}}
\newcommand{\Ec}{E_{\rm conf}}

\newcommand{\Es}{E^{\rm STF}}

\newcommand{\vph}{\varphi^{(m)}_{\rm eff}(z)}
\newcommand{\eins}{\psi^{(1)}}
\newcommand{\zwei}{\psi^{(2)}}

\newcommand{\Ems}{E^{\rm MSTF}}
\newcommand{\xpp}{\x_\perp}
\newcommand{\ypp}{\y_\perp}
\newcommand{\dv}{\frac{dzdp}{2\pi}}
\newcommand{\wt}{\widetilde}
\newcommand{\ol}{\overline}
\newcommand{\ps}{\phi^{\rm STF}}
\newcommand{\psm}{\phi^{\rm MSTF}}
\newcommand{\Ve}{V^{\rm STF}_{\rm eff}}
\newcommand{\bsigma}{\mathord{\hbox{\boldmath $\sigma$}}}
\date{\small\version}

\begin{document}
\markboth{\scriptsize{H \version}}{\scriptsize{H \version}}
\title{\bf{Semiclassics in the lowest Landau band}}
\author{\vspace{5pt} Christian Hainzl\footnote{
current address: Institut f\"ur Mathematik, LMU-M\"unchen, Theresienstr. 39, D-80333 M\"unchen,
E-Mail: \texttt{hainzl@rz.mathematik.uni-muenchen.de}}\\
\vspace{-4pt}\small{Institut f\"ur Theoretische Physik, Universit\"at Wien}\\
\small{Boltzmanngasse 5, A-1090 Vienna, Austria}}

\maketitle

\begin{abstract}
This paper deals with the comparison between the strong Thomas-Fermi
theory and the quantum mechanical ground state energy of a large atom
confined to lowest Landau band wave functions. Using the tools of microlocal
 semiclassical spectral asymptotics we derive precise error
estimates. The approach presented in this paper
suggests the definition
of a modified strong Thomas-Fermi functional,
where the main modification consists in replacing the integration over the
variables
perpendicular to the magnetic field by an expansion in angular momentum
eigenfunctions. The resulting DSTF theory is studied in detail in the second
part of the paper.
\end{abstract}




\section{Introduction}
In this paper we study  semiclassical
theories describing the ground state energies of heavy atoms in
strong homogeneous magnetic fields, where additionally the electrons are
confined to the lowest Landau band.

An atom with $N$
electrons of charge $-e$ and mass $m_e$ and nuclear charge $Ze$ is
described by the nonrelativistic Pauli Hamiltonian
operator
\beq\label{PH}
H_{N}  =  \sum_{1 \leq j \leq N} \left\{ ((-i\nabla^{(j)} + {\bf
A}(x_{j}))\cdot {\bf \sigma}^{j})^{2} - \frac{Z}{|x_{j}|} \right\}
+  \sum_{1 \leq i<j \leq N} \frac{1}{|x_{i} - x_{j}|},
\eeq
acting on the Hilbertspace $\bigwedge_{1 \leq j \leq N} L^{2}({\R}^{3},
{\C}^{2})$ of electrons. The units are chosen such that
$\hbar=2m_{e}=e=1$. The magnetic field is  ${\bf {B}} =
(0,0,B)$, with vector potential ${\bf {A}} =
\frac{1}{2}B(-x_2,x_1,0)$, where $B$ is the magnitude of the field
in units of $B_0 = \frac{m_{e}^{2}e^{3}c}{\hbar^{3}} = 2.35 \cdot
10^{9} $Gauss, the field strength for which the cyclotron radius
$l_{B} = (\hbar c/(eB))^{1/2}$ is equal to the Bohr radius $a_{0}
= \hbar^{2}/(m_{e}e^{2})$. The ground state energy is
\beq
\label{gse} E^{\rm Q}(N,Z,B) = \inf\{ (\psi,H_{N}\psi): \psi \in
{\rm domain} \,\, H_N, (\psi,\psi) = 1\}.
\eeq
Recall that the spectrum of the free Pauli Hamiltonian on
$\Ll^{2}(\R^3;\C^2)$ for one electron in the magnetic field ${\bf B}$,
\beq
H_\A =  \left[\bsigma\cdot(-i\nabla + {\bf A}(\x))\right]^{2},
\eeq
is given by
\beq
p_z^2 + 2\nu B  \qquad \nu = 0,1,2,...,\quad p_z\in\R.
\eeq
The projector $\Pi_0$ onto the lowest Landau band, $\nu = 0$, is
represented by the kernel
\begin{equation}
\label{pk} \Pi_{0}(\x,\x') =
\frac{B}{2\pi}\exp\left\{\frac{i}{2}(\xpp \times \xpp') \cdot {\bf
B} - \frac{1}{4}(\xpp - \xpp')^{2}B\right\}\delta(z -
z')P_\downarrow,
\end{equation}
where $\xpp$ and $z$ are the components of $\x$ perpendicular and
parallel to the magnetic field, and $P_\downarrow$ denotes the
projection onto the spin-down component.

In this paper we are especially interested in the ground state energy,
\beq\label{econf}
\Ec^{\rm Q}(N,Z,B) = \inf_{\parallel \psi \parallel =1} (\psi,\Pi_0^N
H_N \Pi_0^N \psi),
\eeq
where $\Pi_0^N$ denotes the $N$-th tensorial power of $\Pi_0$. 
Lieb, Solovej and Yngvason pointed out
 that  for $B \gg Z^{4/3}$ the electrons are to the leading
order confined to the lowest Landau band, which is expressed by the
following theorem.

\begin{thm}\label{limqstf}
(\cite{LSY1}, Theorem 1.2) For any fixed $\lambda=N/Z$ there is
a $\delta(x)$ with $\delta(x)\to 0$ as $x\to\infty$ such that
\beq
\Ec^{\rm Q}\geq E^{\rm Q}\geq \Ec^{\rm
Q}\left(1+\delta(B/Z^{4/3})\right).
\eeq
\end{thm}

The energy 
(\ref{econf}) can be approximated by means of the STF-functional (Strong Thomas-Fermi)
\beq
\label {stff} {\mathcal{E}} ^{\rm STF}[\rho ] =
\frac{4\pi^{4}}{3B^{2}} \int \rho^{3} - \int V\rho + D(\rho ,\rho),
\eeq
$V(x) = Z/|x|$ and $D(\rho,\rho) =
\frac{1}{2}(\rho,|x|^{-1}\ast\rho)$.
In \cite{LSY2} it is  shown that $E^{\rm Q}/E^{\rm STF} \to 1$ if $Z\to
\infty$, $B/Z^3 \to 0$ and $B/Z^{4/3}\to \infty$, where
\beq\label{stf}
E^{\rm STF}(N,Z,B) = \inf\{{\mathcal {E}}^{\rm STF}[\rho] | \rho
\geq 0, \rho \in D^{\rm STF}, \int\rho \leq N\},
\eeq
with an appropriately chosen domain $D^{\rm STF}$.
Combined with Theorem \ref{limqstf} this implies Theorem
\ref{econftostf}.
\begin{thm}\label{econftostf}
(\cite{LSY2})
If $Z \to \infty$ with $N/Z$
fixed, $B/Z^{3} \to 0$ and $B/Z^{4/3} \to \infty$, then
\begin{equation}
\Ec^{\rm Q}(N,Z,B)/E^{\rm STF}(N,Z,B) \rightarrow 1.
\end{equation}
\end{thm}

\subsection{Comparing the STF energy with the QM ground state energy in the lowest Landau band}

In this paper we not only want to give a direct proof of Theorem
\ref{econftostf}, but we want to derive precise error estimates.
In this respect our procedure is related to \cite{IS} and
\cite{Ivrii1}. Our main theorem is the following:
\begin{thm}\label{qstf}
Let $N \sim Z$ and $Z^{4/3} \leq B\leq Z^3$. Then
\beq\label{eqstf}
|\Ec^{\rm Q} (N,Z,B) - \Es (N,Z,B)| \leq C B^{4/5}Z^{3/5}
\eeq
for an appropriate constant $C$.
\end{thm}

\begin{rem}
In \cite{Ivrii1} Ivrii estimated the difference between the full quantum
mechanical energy $E^{\rm Q}$ and the MTF energy, which is given by minimizing
the MTF
functional (cf. \cite{LSY2}), where all Landau levels are taken into account.
The main theorem in \cite{Ivrii1} reads:
\begin{thm}(\cite{Ivrii1} Theorem 0.2)\label{Ivrii0}
Let $ B \leq  Z^{3}$ and $N \sim Z$, then
\begin{equation}\label{Ivrii}
|E^{ Q}(N,Z,B) - E^{\rm MTF}(N,Z,B) - \frac{1}{4} Z^{2}| \leq R_{1} + R_{2},
\end{equation}
with
\begin{equation}\label{Ivrii2}
R_{1} = CZ^{4/3}(N + B)^{1/3} \quad and \quad R_{2} = CZ^{3/5}B^{4/5}.
\end{equation}
\end{thm}
\begin{rem}
Although true for all $B \leq Z^3$, it should be noted that only
for $B < Z^{7/4}$ are the error terms $R_1$ and $R_2$ smaller than
the Scott term $\frac 14 Z^2$.
\end{rem}

One of the main difficulties Ivrii has to cope with in the proof
of Theorem \ref{Ivrii0} is the fact that the self-consistent MTF
potential  is not  smooth, because it includes all Landau levels.
So he has to create an approximating $C^\infty$ potential in order
to apply the tools of microlocal  semiclassical spectral
asymptotics. Fortunately, in our
case of Theorem \ref{qstf} we need not care about such problems,
since the STF potential (see (\ref{023})) has all required
properties for semiclassical spectral asymptotics.

Moreover, in Theorem \ref{Ivrii0} Ivrii already captures  (\ref{eqstf})
on the region where only the lowest Landau band is occupied, i.e on
$\{\x| |\x| \geq C_0 Z/B\}$ with a large constant $C_0$.
\end{rem}

We mention some important steps  of the proof of Theorem \ref{qstf}.\\
Let $\ps $ denote the effective STF potential (for simplicity think  of the
neutral case $N=Z$)
\beq\label{023}
\ps (x) =Z |x|^{-1}- \rs\ast |x|^{-1},
\eeq
where $\rs$ is the minimizer of the STF functional (\ref{stff}).
We will see in Section 2.1 that the main contribution of (\ref{eqstf}) is given
by
\beq\label{ee1}
\left| \Tr [\Pi_0 ( H_\A - \ps(\x))\Pi_0]_- - \frac B{2\pi}
\int_{\R^3} \int_\R \frac{dp d\x}{2\pi} [p^2 - \ps(\x)]_-\right|,
\eeq
with $[t]_- = \min\{0,t\}$. Recall
\beq
\Es =  \frac B{2\pi} \int_{\R^3}
\int_\R \frac{dp d\x}{2\pi} [p^2 - \ps(\x)]_- - D(\rs,\rs).
\eeq
With the decomposition
$\Ll^2(\R^3,d\x;\mathbb{C}^2)=\Ll^2(\R^2,d\xperp)\otimes \Ll^2(\R,dz)\otimes\mathbb{C}^2$
the projector $\Pi_0$ can be written as
\beq\label{dec}
\Pi_0=\sum_{m\geq 0}|\phi_m\rangle\langle\phi_m|\otimes{1}\otimes P_\downarrow,
\eeq
where $\phi_m$ denotes the function in the
lowest Landau band with angular momentum $-m\leq 0$,
i.e., using polar
coordinates $(r,\varphi)$,
\beq\label{amf}
\phi_m(\xperp)=\sqrt\frac B{2\pi}\frac 1{\sqrt{m !}}\left(\frac
{Br^2}{2}\right)^{m/2}e^{-i m \varphi}e^{-B r^2/4}.
\eeq
Using this and $H_{\bf A} \Phi_m =0$, we can write
\beq\label{pi0}
\Pi_{0}H_{\A}\Pi_{0} = \sum_{m\geq 0}|\phi_m\rangle\langle\phi_m |
\otimes (-\partial_z^2)\otimes P_\downarrow.
\eeq
By means of the above decompositions one gets the relation
(cf. \cite{H1} Theorem 3.13)
\beq\label{tre}
\Tr [\Pi_0 (-\partial_z^2 - \ps(\x))\Pi_0]_- = \sum_m \T
[-\partial_z^2 - \ps_m (z)]_- ,
\eeq
with
\beq
\ps_m(z) = \int d\xpp \ps (\x) |\phi_m(\xpp)|^2.
\eeq
Next we multiply the $m$-th term of the right hand side of
(\ref{tre}) with $\frac{B}{2\pi}\chi_m (\x_\perp)$, where
\beq
\chi_m (\x_\perp) = \left\{
\begin{array}{cc} 1 & {\rm for} \,\,\, \sqrt{2m/B} \leq |\xpp|\leq
\sqrt{2(m+1)/B} \\
0 & {\rm otherwise},  \end{array} \right.
\eeq
and integrate over $\x_\perp$, which is just
an identity operation. Since we are allowed to put the sum into
the trace as well into the $[]_-$ bracket we arrive at
\beq\label{intre}
\Tr [\Pi_0 [-\partial_z^2 - \ps(\x)]\Pi_0]_- =
\frac B{2\pi} \int d\xpp \T
[-\partial_z^2 - \widetilde \ps (\x)]_- ,
\eeq
with
\beq\label{tps}
\widetilde \ps (\x) = \sum_m \chi_m (\x_\perp) \ps_m(z).
\eeq
The Equation (\ref{intre}) follows from the fact that the terms
$\chi_m(\xpp)\ps_m(z)$ (\ref{tps}) have disjoint supports.

Hence, (\ref{ee1}) can be written as
\beq\label{e2}
\left| \frac B{2\pi} \int_{\R^2} d\xpp \left (
 \T [ -\partial_z^2 - \widetilde\ps(\x)]_-  -  \int_{\R^2} \frac{dzdp}{2\pi}
 [p^2 -
\ps(\x)]_-\right ) \right|.
\eeq
We shall estimate (\ref{e2}) by splitting into the following two
terms:
\beq\label{e4}
\left| \frac B{2\pi} \int_{\R^2} d\xpp \left (
 \T [ -\partial_z^2 - \widetilde \ps(\x)]_-  -  \int_{\R^2}
 \frac{dzdp}{2\pi} [p^2 - \widetilde
\ps(\x)]_-\right ) \right|
\eeq
and
\beq
\label{5} \left| \frac B{2\pi} \int_{\R^2}
d\xpp  \left ( \int_\R \frac{dzdp}{2\pi} [p^2 -
\widetilde \ps(\x)]_- - \int_{\R^2} \frac{dzdp}{2\pi} [p^2 -
\ps(\x)]_- \right)\right|.
\eeq

\subsection{Modified STF functionals}

{}From Equations (\ref{tre}) and (\ref{intre}) it is apparent that
the STF energy (\ref{stf}) is not the most natural semiclassical
approximation of $\Ec^{\rm Q}$. As already argued in \cite{H1},
(\ref{tre}) suggests the definition of a functional, where the
integration over $\xpp$, the variables orthogonal to the magnetic
field, is replaced by an expansion in angular momentum
eigenfunctions in the lowest Landau band. This leads to a {\it
discrete STF
functional} (DSTF) depending on a sequence of one-dimensional
densities $\rho = (\rho_n)_{n \in \N_0}$, i.e.
\beq\label{idstff}
\E[\rho]=\sum_{m\in\N_0}\left(\kappa\int \rho_m(z)^3 - Z\int
V_m(z)\rho_m(z)dz\right)+\overline D(\rho,\rho),
\eeq
where $\kappa = \pi^2 /3$,
\beq
\overline D(\rho,\rho)= \frac 12 \sum_{m, n}\int
V_{m,n}(z-z')\rho_m(z)\rho_n(z') d z  d z',
\eeq
and the potentials $V_m$ and $V_{m,n}$ are given by
\beqa\nonumber
V_m(z)&=& \int \frac 1{|\x|}|\phi_m(\xperp)|^2 d\xperp, \\ \label{pot}
V_{m,n}(z-z') &=& \int \frac {|\phi_m(\xperp)|^2 |\phi_n(\xperp
')|^2}{|\x-\x'|} d\xperp d\xperp'.
\eeqa

An equivalent functional, depending on a three-dimensional density $\rho$,
can be obtained as in (\ref{intre}),
if in STF theory the Coulomb potential is replaced by
\beq
\wt {|\x|^{-1}} = \sum_m \chi_m (\x_\perp) \int_{\R^2} d\xpp
|\x|^{-1} \phi_m (\xpp).
\eeq
The resulting {\it modified} STF functional is given by
\beq
\label{imstff} {\mathcal{E}} ^{\rm MSTF}[\rho ] =
\frac{4\pi^{4}}{3B^{2}} \int d\x \rho(\x)^3 - \int d\x
\wt {|\x|^{-1}} \rho(\x) + \wt D(\rho ,\rho
),
\eeq
with
\beq
\wt D(\rho,\rho)= \frac 12 \sum_{m, n}\int d\x d\y
V_{m,n}(z-z')\chi_m(\xpp)\chi_n(\ypp) \rho(\x) \rho(\y).
\eeq
Let $\rho^{\rm DSTF}=(\rd_m(z))_m$ and $\rho^{\rm MSTF}=\rmd(\x)$ be the
ground state
densities of (\ref{idstff}) respectively (\ref{imstff})
corresponding to a fixed particle number $N$. Then the
relationship between the densities reads
\beq
\rho^{\rm MSTF}(\x) = \frac B{2\pi} \sum_m \rho_m^{\rm
DSTF}(z)\chi_m (\xpp).
\eeq
Furthermore the energies are equal,
\beq
E^{\rm MSTF} (N,Z,B) = E^{\rm DSTF}(N,Z,B).
\eeq

Since a short computation shows that for $ B\geq Z^{4/3}$
the difference between the D(M)STF energy and the STF
energy is smaller than $B^{4/5}Z^{3/5}$
the estimate (\ref{eqstf}) with STF replaced by D(M)STF
immediately follows for this region
($B\geq Z^{4/3}$).
\begin{thm}
Let $ Z^{4/3} \leq B \leq Z^3$ and $N \sim Z$. Then
\beq
|\Ec^{\rm Q}(N,Z,B) - E^{\rm D(M)STF}(N,Z,B)| \leq CB^{4/5}Z^{3/5}.
\eeq
\end{thm}

\section{Proof of Theorem \ref{qstf}}

\subsection{Derivation of lower and upper bounds to $\Ec^{\rm Q}$}

First of all, recall that the TF equation satisfied by the
minimizer of (\ref{stff}) under the constraint $\int \rho =N$
is
\beq\label{stfe}
\kappa (\rs)^2/B^2 = [Z|\x|^{-1} - \rs + \nu]_+
= [\ps + \nu]_+ ,
\eeq
where $\nu = \nu(N)$ is the chemical potential corresponding to the
electron number $N$. Using (\ref{stfe}) one sees
that the STF energy (\ref{stf}) can be written as
\beq\label{edefstf}
\Es =  \frac B{2\pi} \int_{\R^2}d\xpp
\int_\R \dv [p^2 - \ps(\x) - \nu]_- + \nu N- D(\rs,\rs).
\eeq
This expression will be used for deriving upper and lower bounds
to the quantum mechanical ground state energy $\Ec^{\rm Q}$.

{\it Lower bound}:\\
Let $\psi$ denote a ground state wave function of $H_N$, then we
can write $\Ec^{\rm Q}$ as
\beqa\nonumber
\Ec^{\rm Q} &=& (\psi,\Pi_0^NH_N \Pi_0^N \psi)\\ \nonumber
&=&
\sum_{i
=1}^{N}(\psi,\Pi_0^N[H_A(\x_i) - Z|\x_i|^{-1}+\rs\ast|\x_i|^{-1} -
\nu]\Pi_0^N\psi)\\ \label{e21}
&&
+N\nu - 2D(\rho_\psi,\rs) + \sum_{i<j} (\psi,|\x_i - \x_j|^{-1}\psi),
\eeqa
where we have added and subtracted the term $\rs\ast |\x|^{-1}
-\nu$ and used the definition
\begin{equation}
\rho_\psi (\x) = N\sum_{s^{i}}\int |\psi(\x,x_{2},..,x_{N};s^{1},..,s^{N})|^{2}
dx_{2}..dx_{N}.
\end{equation}
By means of the Lieb-Oxford inequality \cite{LO}
\beq\label{lo}
\sum_{i<j} (\psi,|\x_i - \x_j|^{-1}\psi) \geq D(\rho_\psi,
\rho_\psi) - 1.68\int \rho_\psi^{4/3},
\eeq
(\ref{e21}) can be bounded from below by
\beq
\Ec^{\rm Q} \geq \Tr [\Pi_0(H_\A - \ps - \nu)\Pi_0]_- - D(\rs,\rs) +\nu N -
1.68\int\rho_\psi^{4/3},
\eeq
where we have used that
\beq\label{d}
D(\rho_\psi - \rs, \rho_\psi -\rs) \geq 0.
\eeq
Furthermore  by (\ref{intre}) and (\ref{edefstf}) we get
\beq\label{lb}
\Ec^{\rm Q} \geq \Es - {\mathcal{R}}_1 - 1.68\int\rho_\psi^{4/3},
\eeq
with
\beqa\nonumber &&
{\mathcal{R}}_1 =\left| \frac B{2\pi} \int_{\R^2} d\xpp \left (
 \T [ -\partial_z^2 - \widetilde\ps(\x) - \nu]_-  -\qquad\qquad \right.\right.\\
\label{e22}&&
\qquad\qquad\qquad \left. \left. -\int_{\R^2} \frac{dzdp}{2\pi}
 [p^2 -
\ps(\x) -\nu]_-\right ) \right| .
\eeqa
Since $\psi$ is a ground state wave function, or at least an
approximate ground state wave function, we can estimate (cf.
\cite{LSY1} (8.5))
\beqa
\nonumber
\int\rho_\psi^{4/3} &\leq& (\int \rho_\psi^3)^{1/6}
(\int\rho_\psi)^{5/6} \leq {\rm const.} (B^2 |\Es|)^{1/6}N^{5/6}\\
&\leq& {\rm const.} Z^{1/5} N^{14/15} B^{2/5} \leq CB^{4/5}
Z^{3/5},
\eeqa
using $N \sim Z$ and $B \geq Z^{4/3}$.

{\it Upper bound}:\\
For every fixed integer $N$ and a normalized $N$-particle wave function $\psi$
we have
\beqa\nonumber
\Ec^{\rm Q} &\leq& (\psi,\Pi_0^NH_N \Pi_0^N\psi)\\ \nonumber
&=&
\sum_{i
=1}^{N}(\psi,\Pi_0^N[H_A(\x_i) - Z|\x_i|^{-1}+\rs\ast|\x_i|^{-1} -
\nu]\Pi_0^N\psi)\\ \label{e23}
&&
+N\nu - 2D(\rho_\psi,\rs) + \sum_{i<j} (\psi,|\x_i -
\x_j|^{-1}\psi).
\eeqa
We set
\beq\label{psi}
\psi = \frac{1}{\sqrt{N!}} \phi_{1} \wedge ... \wedge
\phi_{N},
\eeq
where $\phi_i$ is the eigenvector corresponding to the $i$-th lowest
eigenvalue $\lambda_i$ of the one-particle operator
\beq\label{op}
\Pi_0 (H_A - Z|\x|^{-1} + \rs\ast|\x|^{-1})\Pi_0.
\eeq
By means of the decomposition
\beq
\Tr [\Pi_0(H_A(\x) - \ps(\x) -
\nu)\Pi_0]_- = \sum_i [\lambda_i -\nu]_- = \sum_{i=1}^N (\lambda_i
-\nu) + \sum_{\lambda_N < \lambda_i <\nu}  (\lambda_i -\nu),
\eeq
and the equation (\ref{d}) we can
estimate (\ref{e23}) as
\beqa\nonumber
\Ec^{\rm Q} &\leq& \Tr[\Pi_0 (H_A - \ps  -
\nu)\Pi_0]_-  + \nu N - D(\rs,\rs)\\&& +D(\rho_\psi - \rs,\rho_\psi- \rs)
-\sum_{\lambda_N < \lambda_i < \nu} (\lambda_i -
\nu),
\eeqa
which implies
\beq\label{ub}
\Ec^{\rm Q} \leq {\mathcal{R}}_1 + R_2 + R_3,
\eeq
with
\beqa
&{{R}}_2& =
-\sum_{\lambda_N < \lambda_i <\nu} (\lambda_i -
\nu),\\
 &{{R}}_3& =D(\rho_\psi - \rs,\rho_\psi- \rs).
\eeqa
Since it is difficult to tackle directly the term $R_2$,
we estimate $R_2$ by $|\lambda_N - \nu|$ multiplied by the number
of eigenvalues of the operator (\ref{op}) between $\lambda_N$ and
$\nu$.
We know
\beq
\frac B{2\pi} \int d\xpp \int \dv
\Theta_-(p^2 -  \ps (\x)- \nu) = \int \rs = N,
\eeq
with
\beq
\Theta_-(t) = \left\{ \begin{array}{cc}
1 & {\rm for} \,\,\, t\leq 0\\
0 & {\rm otherwise}.
\end{array} \right.
\eeq
So the number of eigenvalues of (\ref{op}) between $\lambda_N$ and $\nu$
can be expressed by
\beq\label{nmeigv}
\Tr \Theta_-(\Pi_0(-\partial_z^2 - \ps(\x)- \nu)\Pi_0) -
\frac B{2\pi} \int d\xpp \int \dv
\Theta_-(p^2 -  \ps (\x)- \nu).
\eeq
Mimicking the derivation of (\ref{intre}), with
$[t]_-$ being  replaced by $\Theta_-(t)$, leads to
\beq
(\ref{nmeigv}) =
\frac B{2\pi} \int_{\R^2} d\xpp \left (
\T \Theta_- ( \wt H_{\xpp}-\nu)  -  \int \frac{dzdp}{2\pi} \Theta_-(h_{\xpp} -\nu)
\right ),
\eeq
where we have defined
\beq\label{th}
\wt H_{\xpp} =-\partial_z^2 - \widetilde\ps(\x) \quad {\rm and}
\quad h_{\xpp} =p^2 - \ps(\x).
\eeq
Hence, instead of $R_2$ we estimate the error term
\beq
{\mathcal{R}}_2 = |\lambda_N - \nu|
\frac B{2\pi} \int_{\R^2} d\xpp \left (
\T \Theta_- (\wt H_{\xpp} -\nu)  -  \int \frac{dzdp}{2\pi} \Theta_-(h_{\xpp}-\nu)
\right).
\eeq

We introduce the notation
\beq
e(\wt H_{\xpp},\mu) = \Theta_-(\wt H_{\xpp} -\mu),
\eeq
the projector of the operator $\wt H_{\xpp}$ onto
the eigenspace corresponding to the eigenvalues smaller or equal
to
$\mu$. Let $\psi$ be given by (\ref{psi}), then we have
\beq
\rho_\psi(\x) = \frac B{2\pi}e(z,z;\wt
H_{\xpp},\lambda_N),
\eeq
whereas the semiclassical density $\rs$ can be written as
\beq
\rs(\x) = \frac{B}{2\pi} \int_\R dp\Theta_-(h_{\xpp}-\nu)=
 \frac{B}{2\pi}[\ps(\x) + \nu]_+^{1/2}.
\eeq
Furthermore we introduce the auxiliary density
\beq\label{br}
\bar \rho (\x) = \frac{B}{2\pi} [\ps(\x) + \lambda_N]_+^{1/2}.
\eeq
In order to  bound the error term $R_3$ from above, we combine the two terms
\beqa
&{\mathcal{R}}_3& = D(\rho_\psi - \bar \rho,\rho_\psi -
\bar\rho),\\
&{\mathcal{R}}_4& = D(\rs - \bar \rho,\rs - \bar\rho),
\eeqa
which are easier to  handle than $R_3$ alone. Observe that by convexity one
has $R_3 \leq 2\Rm_3 + 2\Rm_4$.
In the next sections we will separately have to carry out
the estimations
\beq
{ \mathcal{R}}_i \leq CB^{4/5}Z^{3/5} \quad \forall i = 1,..,4.
\eeq

\subsection{Methods used in the proof}

The methods used here in order to estimate the error terms ${
\mathcal{R}}_{1-4}$, have been established in \cite{IS} and
\cite{Ivrii2}. For sake of better understanding we will state here
the most important theorem, which we use throughout this section.

Consider the Schr\"odinger operator
\beq
H = -\frac 12 \Delta - \phi(\x) \quad {\rm on} \,\,\, \R^d.
\eeq
Its symbol is denoted by
\beq
\bar h = \frac 12 p^2 - \phi(\x).
\eeq
In addition there are the following conditions imposed on the real
potential $\phi$:

There are Lipschitz functions $l(\x) > 0$ and $f(\x) >0$, such
that:
\beqa\label{c1}
&(i)& |\nabla l(\x)| \leq M,\\
&(ii)& cf(\y) \leq f(\x) \leq Cf(\y),\\
\label{c3}
&(iii)& |\partial^{\nu} \phi(\x)| \leq C_\nu f(\x)^2
l(\x)^{-|\nu|} \,\,\, \forall \nu \in \N^d
\eeqa
Under these assumptions Ivrii and Sigal have proved the following
theorem:
\begin{thm}\label{IS1}
(\cite{IS} Theorem 7.1.)\\ Assume conditions $(i) - (iii)$ are
obeyed and let $\psi$ be smooth and obey $|\partial^\nu
\psi(\x)|\leq C_\nu l(\x)^{-|\nu|}$ for any $\nu$. Let
$g_s(\lambda) = [-\lambda]_+^s$ for some $s \in [0,1]$. Then
\beqa
&&\left|{\rm Tr}(\psi g_s(H)) - \int  \frac{dx dp}{2\pi}\psi
g_s(\bar h)\right|\qquad \qquad\\ &&\qquad \quad\quad \quad \leq C\int_{{\rm
supp}\psi} d\x \max \left[\left(\frac 1{f(\x)l(\x)}\right)^{\alpha
- s - d} , 1\right] l(\x)^{-2s -d},
\eeqa
Here $\alpha =1$ if either $d\geq 2$ or $d=1$ and $\phi$ obeys
\beq\label{d1c}
|\phi(\x)| + l(\x) |\nabla\phi(\x)| \geq \eps f(\x)^2
\eeq
on $\{\x \, | l(\x)f(\x) \geq 1\}$, with some $\eps > 0$,
and $\alpha = 1/2$ otherwise.
\end{thm}

The most important tools for the proof of Theorem \ref{IS1}
are multiscale analysis and semiclassical spectral asymptotics.
First of all, the domain, i.e. the support of $\psi$, is covered by
a countable number of balls. Then on each of these balls
$B(\y,l(\y))$,
the operator $H$ is transformed into
\beq\label{khv}
K_h = f^{-2} U(l) H U(l)^{-1} = -\frac{h^2}2 \Delta - V(\x),
\eeq
with $h=l(\y)^{-1}f(\y)^{-1}$,
by means of a unitary scaling transformation $U(l)$,
which maps  the ball $B(\y,l(\y))$ into $B(0,1)$.
Next Theorem \cite{IS} 6.1, which we state below, is applied to the sum of the negative eigenvalues of
$g_s(K_h)$. After rescaling and summing over all balls one arrives at Theorem
\ref{IS1}.

The  symbol of $K_h$ is given by
\beq
k(\x,p)= p^2/2 - V(\x).
\eeq
If furthermore all derivatives of $V$ are bounded by a
constant, i.e.
\beq\label{vicond}
|\partial^\nu V (\x)| \leq C_\nu \quad {\rm on} \,\,
B(0,2)\,\, \forall i,
\eeq
then the following theorem is valid:
\begin{thm}\label{IS2}
Let $\psi \in C^\infty_0 (B(0,1))$ and let $0$ be a regular value of
the
function $k_0$ restricted to ${\rm supp}\psi\times \R^d$. Then for
$h\leq 1$
\beq
{\rm Tr}(\psi g_s(K_h)) = h^{-d}\int \frac{dxdp}{2\pi} \psi g_s(k
(x,p)) + O(h^{s+1-d}).
\eeq
\end{thm}
Assume next that the potential $V$ can be written as
\beq
V(\x) = V_0(\x) + h\bar V(\x,h),
\eeq
such that the principal symbol of $K_h$ reads
$k_0(\x,p) = p^2/2 - V_0(\x)$.
Then, if additionally  $\bar V$ fulfills (\ref{vicond}) (uniformly in $h$),
Theorem \ref{IS2} remains valid for $s=0$ with $k$ replaced by the
principal symbol $k_0$. (cf. \cite{Ivrii2} Theorem 4.5.3)
\begin{thm}\label{IS3}
Let $\psi \in C^\infty_0 (B(0,1))$ and let $0$ be a regular value of
the
function $k_0$ restricted to ${\rm supp}\psi\times \R^d$. Then for
$h\leq 1$
\beq
{\rm Tr}(\psi g_0(K_h)) = h^{-d}\int \frac{dxdp}{2\pi} \psi g_0(k_0
(x,p)) + O(h^{1-d}).
\eeq
\end{thm}

\subsection{Relationship between the potentials
$\ps$ and $\wt \ps$.}

Next we collect some information about the potentials $\ps$ and
$\wt \ps$.

The scaling functions, which we will use in order to apply
Theorem \ref{IS1} to the operator $H_{\xpp}$, have to be chosen such that the
conditions (\ref{c1})-(\ref{c3}) hold for the potentials $\ps$ and $\wt \ps$,
at least away from the origin.

Since we will see that $\ps$ behaves like $Z|\x|^{-1}$
for $|\x| \leq r_S$, the edge of the STF atom, it is thus natural
to define
\beq\label{scalf}
l(\x) = ({\rm const.})|\x| \quad {\rm and} \quad
f(\x)^2 =Z|\x|^{-1}.
\eeq

If we denote the effective STF potential as
\beq
\Ve (\x) = \ps(\x) +\nu,
\eeq
then  the following Lemma is valid:
\begin{lem}\label{lem1}
(i) \, (\cite{LSY2}) The density $\rs(\x)$ as well as $[\Ve(\x)]_+$ have compact
support with  radius $r_S \leq 3.3 \pi^2 Z^{1/5}B^{-2/5}$.\\
(ii)\, For fixed but arbitrary  $\xpp $, $ \ps(\x)$ and $ \wt \ps (\x)$  are
$ \in C^\infty(\R
\setminus {0})$ as a function of
$z$ and
\beq\label{pe}
|\partial_z^{\nu} \ps(\x)|\leq C_\nu f(\x)^2 l(\x)^{-\nu}, \qquad
|\partial_z^{\nu}\wt \ps (\x)|\leq C'_\nu f(\x)^2 l(\x)^{-\nu},
\eeq
for all $\nu \in \N$ and $\x \in \R^3$.
\end{lem}

\begin{proof}
$(ii)$ The $C^\infty $ property follows from the TF equation
(\ref{stfe}) and the definition (\ref{tps}). Equation (\ref{pe}) follows
from (\ref{stfe}) and (\ref{tps}) and the fact, that
\beq
|\partial_z^\nu Z|\x|^{-1}| \leq f(\x)^2|\x|^{-\nu} \frac
z{|\x|} \leq f(\x)^2l(\x)^{-\nu}.
\eeq
The proof of $(i)$ is given in \cite{LSY2} Theorem 4.11.
\end{proof}
\begin{rem}
The estimates (\ref{pe}) seem to be very crude, especially in the
vicinity of $r_S$, but nevertheless they are good enough to
provide precise error estimates. Next
let us try to get an idea how $\Ve$ behaves
in the vicinity of the radius $r_S$.
We consider the neutral case $N=Z$.
Since $\Ve$ is spherical symmetric, we can make the ansatz
$\Ve (\x) = \chi(r) /r$, $(|\x| = r)$, which leads
by (\ref{stfe}) to
\beq\label{chidiff}
\chi''(r) = B r^{1/2}\chi(r) \quad {\rm and} \quad \chi(0) = Z.
\eeq
Around each point $r_0$, this equation has a solution
$\chi$ that can be expanded in a series of
terms $c_i [r_0 - r]^i $ with $i\geq 4$. In the vicinity of $r_0 = r_S$
we get the approximate solution
\beq\label{veffrs}
\Ve \sim \frac{Z^{1/5} B^{8/5}}{r} [r_S - r]_+^4,
\eeq
which shows that $\Ve $ tends to $0$ as $[r_S - r]_+^4$ as $r \to r_S$.
\end{rem}

Next we fix a point $\y \in \R^3$ and we set $l = l(\y)$ and
$f= f(\y)=Z^{1/2}|\y|^{1/2}$. (We assume that (\ref{c1}) - (\ref{c3}) are
fulfilled
in  $B(\y,l(\y))$. This in our case can be done by defining
$l(\y)$ e.g. as $|\y|/2$.)
Furthermore we define the unitary transformation
\beq
U(l): \psi(\x) \to l^{3/2}\psi(l\x + \y),
\eeq
which maps the ball $B(\y,l(\y))$ to $B(0,1)$ and transforms
the operator $-\partial^2_z - \wt \ps$ into
\beq
- l^{-2} \partial^2_z - \wt \ps(l\x + \y).
\eeq
Introduce the new potential
\beq
\wt W(\x) = f^{-2}\wt \ps(l\x +\y).
\eeq
The resulting operator is related to the original one (\ref{th}) as
\beq
U(l) \wt H_{\xpp} U(l)^{-1} = f^2 \wt K_h,
\eeq
with
\beq
\wt K_h = - h^2 \partial^2_z - \wt W (\x) \quad {\rm and} \quad
h= (lf)^{-1}.
\eeq
If we denote $W(\x) = f^{-2}\ps(l\x+\y)$, then one easily sees that
$\wt W$ can equivalently be defined by applying (\ref{tps}), i.e. the
operation $\wt \cdot$, to $W$, with $B$ replaced by $B' = Bl^2$.
In other words the unitary transformation $U(l)$ scales the
magnetic field strength $B$ to $B' = Bl^2$ and for the difference $\wt W - W$ we get the following Lemma.
\begin{lem}
There exists a function $a(\x,\alpha)$  such that
\beq
\wt W (\x) - W(\x) = \alpha a(\x, \alpha),
\eeq
with $\alpha = B^{-1/2} l^{-1}$ and $a(\x,\alpha)$ fulfills
(\ref{vicond}) uniformly in $\alpha$ for $\alpha \leq 1$.
\end{lem}
\begin{proof}
Since the potential $\ps$ is spherical symmetric and Equation
(\ref{pe}) is fulfilled for derivatives in all directions, we get
for $\x \in B(0,2)$
\beq
|\partial^{\bf \nu} W(\x)| = |f^{-2} \partial^{\bf \nu}\ps(l\x +\y)| \leq
C_{\bf \nu} \quad \mbox{for all}  \quad
{\bf \nu} \in \N^3.
\eeq
Hence, $W(\x)$, together with all derivatives, is bounded above by a
constant on a ball around $\x = 0$. Since the operation $\wt \cdot$ smears the
potential, for every $\x$,
over a region $\sim \alpha$ in the $|\xpp|$-direction
the difference $\wt W - W$ can be expressed by $\alpha$ times a
function $a(\x,\alpha)$ which is bounded by a constant. Since
$\wt{ \partial^n_z W}(\x) = \partial^n_z \wt W(\x)$ the same
argument can be given for all derivatives.
\end{proof}
Let us rewrite the operator $K_h$ in the form
\beq\label{kh}
K_h = - h^2 \partial^2_z - (W(\x) +  \alpha a(\x,\alpha)
).
\eeq
In order to be allowed to apply Theorem \ref{IS3}
to (\ref{kh}), i.e. in order to guarantee that
$p^2 - W(x)$ is the principal symbol of $K_h$, it is necessary, that
\beq
\alpha \leq h \quad \Leftrightarrow \quad B^{-1/2} l^{-1} \leq
l^{-1} f^{-1},
\eeq
which leads to the condition
\beq
|\y| \geq Z/B.
\eeq
Hence, in the sense of Theorem \ref{IS3}, this implies that in the
region $\{ \x |\, |\x| \geq Z/B\}$ Theorem \ref{IS1}, with $s=0$ can
be applied to the error terms ${ \mathcal{R}}_{2-4}$, with $\wt
\ps$ replaced by $\ps$.
Next we decompose $\R^3$ into $\Omega_1 = \{\x| |\x|\leq Z/B\}$
and $\Omega_2 = \{ \x |\, |\x| \geq Z/B\}$ and estimate the error
terms ${ \mathcal{R}}_{1-4}$ on each of these regions separately.

\subsection{Analysis in the region $\Omega_1$}

We first assume that $B < Z^2$. This assumption is made
in order to be sure that $\Omega_1$ is not completely contained
in the non-semiclassical region $\{\x| |\x| \leq 1/Z\}$, where
each term, the quantum mechanical as well as the semiclassical,
has to be estimated separately.
Furthermore let $\eins (\x)$ be supported in $\{ \x| \, 0\leq
|\x| \leq Z/B (1 + \epsilon)\}$ and fulfill $\eins(\x) =1$ in
$\{\x| \, |\x| \leq (Z/B) (1 - \epsilon)\}$, as well as
$|\partial_z^n \eins| \leq C_n l(\x)^{-n}$ for all $n \in
\N$.

With respect to $\Rm_1$ and $\Rm_2$, we in particular have to
estimate the term
\beq\label{s1}
\frac B{2\pi} \int d\xpp \left( \T (\eins g_s(\wt H_{\xpp}-\nu)) -
\int \dv \eins g_s(\hp-\nu)\right),
\eeq
with $\wt H_{\xpp}$ and $\hp$ given by (\ref{th}). Let $\tilde
\hp$ be defined analogously, i.e. $\tilde \hp = p^2 - \wt \ps$.
Adding and subtracting $\int \dv \eins g_s(\tilde \hp-\nu)$ in
(\ref{s1}), we split (\ref{s1}) into
\beq\label{s2}
R^s_1(\eins) =
\frac B{2\pi} \int d\xpp \left( \T (\eins g_s(\wt H_{\xpp}-\nu)) -
\int \dv \eins g_s(\tilde \hp-\nu)\right)
\eeq
and the fully semiclassical part
\beq\label{s3}
R^s_2(\eins) =\frac B{2\pi} \int d\xpp \left (\int \dv \eins g_s(\tilde \hp-\nu)
- \int \dv \eins g_s(\hp-\nu)\right ).
\eeq
Since $l(\x)f(\x) \geq 1$, which is equivalent to $|\x| \geq
Z^{-1}$, is necessary for being able to apply Theorem \ref{IS1},
we have to carry out a corresponding decomposition of $\Omega_1$.
Let $\eins_1 + \eins_2 = \eins$ be a partition of unity on $\Omega_1$, with
\beq
{\rm supp}\, \eins_1 = \{\x|\, |\x| \leq Z^{-1}(1 + \epsilon)\},
\,\,
{\rm supp}\, \eins_2 = \{\x|\,Z^{-1}(1 - \epsilon) \leq |\x| \leq Z/B(1 +
\epsilon)\},
\eeq
and $|\partial_z^{n}\eins_i| \leq C_n l(\x)^{-n}$ for $i= 1,2$, with $C_n$
independent of $Z$ and $B$..

\begin{lem}\label{l11}
With above definitions we have for (\ref{s2})
\beq
R^s_1(\eins)
 \leq C B^{\frac 32 s
-1}Z^{2- \frac 12 s}.
\eeq
\end{lem}
\begin{proof}
On ${\rm supp}\,\eins_2$ we apply Theorem \ref{IS1} to (\ref{s1}) with $\alpha
=1$
and $d=1$, which implies for arbitrary but fixed $\xpp$ (we may set the chemical
potential $\nu= 0$ for simplicity, the computations for arbitrary $\nu$ are essentially the same)
\beq
\left| \T (\eins_2 g_s(\wt H_{\xpp})) -
\int \dv \eins_2 g_s(\tilde \hp) \right| \leq
C\int_{{\rm supp} \eins_2 (\xpp,z)} dz l(\x)^{-1-s} f(\x)^s.
\eeq
Hence, multiplying with $\frac{B}{2\pi}$ and integrating over
$\xpp$ leads to
\beq
R^s_1(\eins) \leq C\frac{B}{2\pi} \int d\x l(\x)^{-1-s} f(\x)^s
\leq B Z^{\frac 12 s} \int_{Z^{-1}}^{Z/B} r^{1-
\frac 32 s} dr \leq  C B^{\frac 32 s
-1}Z^{2-\frac 12 s}.
\eeq
In the case of $r\leq Z^{-1}$ the terms of (\ref{s2})
have to be estimated separately.
The semiclassical part reads
\beq
\frac{B}{2\pi} \int \frac{d\x dp}{2\pi} \eins_1 [\tilde \hp]_-^{1/2 +s}
\leq \frac{B}{2\pi}
\int_{|\x| \leq Z^{-1}} d\x [\wt \ps]_+^{1/2 + s} \leq C B Z^{2s -
2}
\eeq
An analogue estimate one derives for $\frac{B}{2\pi} \int d\xpp \T
(\eins_1 g_s(\wt H_{\xpp}))$ by using  \cite{IS} Lemma 7.9.
\end{proof}

\begin{lem}\label{l12}
For (\ref{s3}) we have
\beq
R^s_2(\eins) \leq   CZ^{s+1/2}B^{s/2-1/4}.
\eeq
\end{lem}
\begin{proof}
Obviously, the main contribution to the magnitude of the semiclassical term
\begin{equation}
R^s_2(\eins) \leq \frac{B}{2\pi}\int_{{\R}^{3}} d\x \psi^{(1)}
\left|[\wt {\phi^{\rm STF}}(\x)]_{+}^{s+1/2}  -
[\phi^{\rm STF}(\x)]_{+}^{s+1/2}\right|
\end{equation}
is produced by the Coulomb singularity, i.e
\beqa
R^s_2(\eins)& \leq &\frac{B}{2\pi}\int_{|\x| \leq B^{-1/2}} d\x
|\phi^{\rm STF}(\x)|_{+}^{s+1/2}  \leq\\
&\leq& CB\int_{|\x| \leq B^{-1/2}} d\x\left(\frac{Z}{r}\right)^{s+1/2}
\leq CZ^{s+1/2}B^{s/2-1/4}.
\eeqa
\end{proof}

Hence, we are ready to carry out the estimate of the error terms $\Rm_{1-4}$,
restricted to $\Omega_1$, which we denote as $\Rm_i (\Omega_1)$.
\begin{prop}
For $Z^{4/3} \leq B \leq Z^3$ one has
\beqa\nonumber
&&\Rm_1(\Omega_1) =
\frac B{2\pi} \int_{\R^2} d\xpp \left (
\T(\eins [  \wt H_{\xpp}-\nu]_-)  -  \int \frac{dzdp}{2\pi}\eins [h_{\xpp}
-\nu]_-\right)\\
&&\qquad \qquad \qquad \qquad \leq C B^{4/5} Z^{3/5}.
\eeqa
\end{prop}
\begin{proof}
This is done by putting together Lemma \ref{l11} and Lemma
\ref{l12} and setting $s=1$.
\end{proof}
Before turning to $\Rm_2(\Omega_1)$ we need a preparing Lemma.

\begin{lem}\label{l13}
Let $\lambda_N$ be the $N$-th eigenvalue of (\ref{op}) and $\nu$
the chemical potential of  (\ref{stfe}) belonging to the electron
number $N$.
Then
\beq
|\lambda_N - \nu | \leq C B^{3/5} Z^{1/5}.
\eeq
\end{lem}
\begin{proof}
We assume now that we have already got the estimate
\beq\label{t11}
\left| \Tr \Theta_-(\Pi_0(-\partial_z^2 - \ps)\Pi_0) -
\frac B{2\pi} \int d\xpp \int \dv
\Theta_- (p^2 -  \ps ) \right| \leq C B^{1/5} Z^{2/5},
\eeq
which till now we have only proven on $\Omega_1$, by
setting $s=0$ in
Lemma \ref{l11} and Lemma \ref{l12}. The missing part will be proved
in Lemma \ref{lem21}.
Since we know by definition
\beq
\frac B{2\pi} \int_{\R^2} d\xpp
\T \Theta_- ( \wt H_{\xpp}-\lambda_N)  = N =  \frac B{2\pi}\int d\x
[\ps (\x) + \nu]_+^{1/2},
\eeq
we get
\beqa
& B^{1/5}Z^{2/5} \geq
\frac B{2\pi} \int d\x \left ( [\ps+ \lambda_N]_+^{1/2}
-[\ps + \nu]_+^{-1/2} \right) \geq&\\
& C |\lambda_N - \nu| B \int_0^{r_S} [\ps + \nu']_+^{-1/2} r^2 dr
\geq C |\lambda_N - \nu| (BZ^{-1/2} r_S^{7/2}),&
\eeqa
for some $\nu' \in [\nu, \lambda_N]$. This implies the statement of the
lemma.
\end{proof}
\begin{prop}
\beqa
&&{\mathcal{R}}_2(\Omega_1) = |\lambda_N - \nu|
\frac B{2\pi}\left| \int_{\R^2} d\xpp \left (
\T(\eins \Theta_- (\wt H_{\xpp} -\nu))  - \right.\right.\\
&&\qquad \qquad \qquad\left. \left.- \int \frac{dzdp}{2\pi}\eins
\Theta_-(h_{\xpp}-\nu)\right
 )\right| \leq  C B^{4/5}Z^{3/5}.
\eeqa
\end{prop}
\begin{proof}
By Lemma \ref{l13} and combining the estimations of Lemmata
\ref{l11} and \ref{l12} with $s=0$.
\end{proof}

\begin{rem}\label{rem1}
We remark here that if one has a partition of unity
$\varphi_1 + \varphi_2 =1$, then the relation
\beq\label{dff}
D(f,f) \leq 2D(f\varphi_1, f\varphi_1) + 2D(f\varphi_2,f\varphi_2)
\eeq
is valid, which one gets by the simple inequality
\beq
D(f\varphi_1-f\varphi_2,f\varphi_1-f\varphi_2) \geq 0.
\eeq
\end{rem}

Remark \ref{rem1}  justifies the notations
$\Rm_{3}(\Omega_i), \Rm_{4}(\Omega_i)$, since (\ref{dff}) means
that we can consider each region $\Omega_i$ separately.
\begin{prop}
For $Z^{4/3} \leq B \leq Z^3$   we get with (\ref{br})
\beq
\Rm_4 (\Omega_1) = D(\eins (\rs - \bar \rho),\eins (\rs - \bar
\rho)) \leq CB^{4/5} Z^{3/5}.
\eeq
\end{prop}
\begin{proof}
Recall
\beq
\rs = \frac B{2\pi} [\ps + \nu]_+^{1/2} \,\,\, {\rm and} \,\,\,
\bar \rho = [\ps + \lambda_N]_+^{1/2}.
\eeq
By the Hardy-Littlewood-Sobolev inequality we derive
\beqa\nonumber
\Rm_4 (\Omega_1) &=& D(\eins (\rs - \bar \rho),\eins (\rs - \bar
\rho)) \leq CB^2 \parallel \eins (\rs - \bar \rho) \parallel_{6/5}^2\\
&\leq& C B^2 |\lambda_N - \nu|^2 \parallel \eins (\ps +
\nu')^{-1/2}\parallel_{6/5}^2 \leq Z^{27/5} B^{-14/5}.
\eeqa
\end{proof}
In the case of $\Rm_3(\Omega_1)$ we proceed as above, namely
introduce the auxiliary density $\tilde \rho (\x) = \frac B{2\pi}
[\wt \ps (\x) + \lambda_N]_+^{1/2}$ and decompose $\Rm_3(\Omega_1)$
by using convexity.
\begin{prop}
For $Z^{4/3} \leq B \leq Z^3$ we have
\beq
\Rm_3(\Omega_1) = D(\eins (\rs -  \rho_\psi),\eins (\rs -
\rho_\psi)) \leq CB^{4/5} Z^{3/5}.
\eeq
\end{prop}
\begin{proof}
By decomposition we have on the one hand the fully semiclassical
and easier to handle part
\beq
\Rm_4 (\Omega_1) = D(\eins (\rs - \tilde \rho),\eins (\rs - \tilde
\rho)) \leq B^2 \parallel \eins (\rs - \tilde \rho)
\parallel_{6/5}^2
\leq
B^{4/5} Z^{3/5}.
\eeq
On the other hand there is the more interesting term
\beq
D(\eins (\rho_\psi - \tilde \rho),\eins (\rho_\psi - \tilde
\rho)).
\eeq
For $r\leq Z^{-1}$ we separately calculate
\beqa\nonumber
&(\frac B{2\pi})^2 D(\eins_1 e(z,z;\wt H_{\xpp},\lambda_N),
\eins_1 e(z,z;\wt H_{\xpp},\lambda_N))&\\
\label{d1}&\qquad \qquad \leq CB^2\parallel
\eins_1 e(z,z;\wt H_{\xpp},\lambda_N)\parallel_{6/5}^2&
\eeqa
and
\beq\label{d2}
(\frac B{2\pi})^2 D(\eins_1 [\wt \ps + \lambda_N]_+^{1/2},
\eins_1 [\wt \ps + \lambda_N]_+^{1/2})
\leq CB^2 \parallel \eins_1 [\wt \ps +
\lambda_N]_+^{1/2}\parallel_{6/5}^2.
\eeq
Whereas (\ref{d2}) can be bounded
by
\beq
CB^2 \parallel \eins_1 [Z/|\x|]^{1/2}\parallel_{6/5}^2 \leq
CB^2/Z^3,
\eeq
(\ref{d1}) can analogously be estimated by \cite{IS} Lemma 10.7, or
\cite{Ivrii1} Proposition 4.3.

The term
\beq
D(\eins_2 (\rho_\psi - \tilde \rho),\eins_2 (\rho_\psi - \tilde
\rho))
\eeq
is a bit more delicate. We can either use \cite{Ivrii1}
Proposition 4.3 or \cite{Ivrii2} Theorem 4.5.4 (i),
which states that for $K_h$, given in (\ref{khv}), with $V$ fulfilling
(\ref{vicond}) and $|V(\x) + \tau| \geq \epsilon$
\beq\label{e1}
\left| e(x,x;K_h,\tau) - h^{-d}\int \Theta_- (k(\x,p) -\tau)
\right|\leq h^{1-d}
\,\, \forall \x \in B(0,1/2).
\eeq
Since $|\lambda_N| \leq B^{3/5}Z^{1/5}$, we get that $|\wt \ps +
\lambda_N| \geq \epsilon f(\x)^2$ in $\Omega_1$.
Hence, we can apply (\ref{e1}) to our case, with $d=1$,
yielding
\beq\label{exx}
| e(z,z;\wt H_{\xpp},\lambda_N) - [\wt \ps(\x) +
\lambda_N]_+^{1/2}|\leq l(\x)^{-1}.
\eeq
The term $l^{-1}$ stems from rescaling $B(0,1) $ to $B(0,l)$.
So,
\beq
D(\eins_2 (\rho_\psi - \tilde \rho),\eins_2 (\rho_\psi - \tilde
\rho)) \leq C B^2 \parallel \eins_2 l(\x)^{-1}\parallel_{6/5}^2
\leq CB^2[Z/B]^3.
\eeq
\end{proof}
In the case $B \geq Z^2$, $Z/B$ is smaller than $1/Z$ and in the
above calculations only the separate terms have to be taken into
account, which yields analogue estimates as above.

\subsection{Analysis in the outer Region $\Omega_2$}

This region has already been treated by Ivrii in \cite{Ivrii1} Section
4.

Recall first that $r_S$ is the radius  of the support of $\ps$, in the
neutral case, and of $[\ps +\nu]_+$ otherwise. In order that  Theorem \ref{IS1}
can be applied $\ps $ and $\wt \ps$ have to fulfill condition (\ref{d1c}). We know that
$\nabla \ps(r_S) = \ps(r_S)=0$. Hence,
we look for a parameter $0<c<1$,
and the concerning radius $cr_S$,
such that the separate quantum mechanical as well as semiclassical
parts of $\Rm_{1-4}$ in $\{\x| |\x| \geq cr_S\}$ do not exceed $CB^{4/5}Z^{3/5}$ and that $\ps$ fulfills
(\ref{d1c}).
The existence of such a $c$ is a  consequence of the
behavior of $\ps$ in the vicinity of $r_S$ (cf. (\ref{veffrs})).
By means of such a parameter $c$ we decompose the outer region
$\Omega_2$ into $\Omega_2^1 \cup \Omega_2^2$ and define a concerning partition
of unity, i.e.
\beq
{\rm supp} \zwei_1 = \{\x|\, [Z/B](1 + \epsilon)\leq |\x| \leq cr_S\},
\,\,
{\rm supp} \zwei_2 = \{\x|\, |\x| \geq cr_S (1 - \epsilon) \},
\eeq
with $\zwei_1 + \zwei_2 = 1$ for $r\geq [Z/B](1+\epsilon)$.
On $\Omega^2_2$, by definition, all terms separately are bounded above
by $ CB^{4/5}Z^{3/5}$ and on $\Omega_2^1$ condition
(\ref{d1c}) is fulfilled for $\eps$ small enough.

Throughout this section we assume $Z^{4/3} \leq B \leq Z^3$.
\begin{prop}\label{r1o2}
\beq
\Rm_{1} (\Omega_2^1) \leq C B^{4/5} Z^{3/5}.
\eeq
\end{prop}
\begin{proof}
First we assume $B < Z^2$.  Applying  Theorem \ref{IS1} with $\alpha =1$
and $d=1$ we get for arbitrary but fixed $\xpp$ (we set $\nu
=0$)
\beq\label{prop33}
\left|\T (\zwei_1 g_1(\wt H_{\xpp})) -
\int \dv \zwei_1 g_1(\tilde \hp)\right| \leq
C\int_{{\rm supp} \zwei_1 (\xpp,z)} dz l(\x)^{-2} f(\x).
\eeq
After multiplying with $B$ and integrating over $\xpp$ we get
\beq\label{2s33}
(\ref{prop33}) \leq C\frac{B}{2\pi} \int d\x l(\x)^{-2} f(\x)
\leq C B Z^{1/2}\int_{Z/B}^{cr_S} r^{-
\frac 12 } dr \leq  C BZ^{1/2}[r_S]^{\frac 12}.
\eeq
In the case of $B > Z^2$ we again have to decompose $\Omega_2^1$,
since $Z/B$ is smaller than $1/Z$. So for fixed but arbitrary $B$,
$\Omega_2^1$ is decomposed with respect to $r=1/Z$.
For $r\leq 1/Z$ we proceed as in the previous section and estimate
each term separately and for $r\geq 1/Z$ we immediately arrive at
(\ref{2s33}).

The pure semiclassical part
\beq\label{sp3}
\frac B{2\pi} \int d\xpp \left (\int \dv \zwei g_1(\tilde \hp-\nu)
- \int \dv \zwei g_1(\hp-\nu)\right )
\eeq
can analogously be estimated as in Lemma \ref{l12}.
\end{proof}
Denote
\beq\label{2s2}
R^0(\zwei_1) =
\frac B{2\pi} \int d\xpp \left( \T (\zwei_1 \Theta_-(\wt H_{\xpp}-\nu)) -
\int \dv \zwei_1 \Theta_-( \hp-\nu)\right).
\eeq
\begin{lem}\label{lem21}
\beq
R^0 (\zwei_1) \leq CB^{\frac 15 }Z^{\frac 25 }.
\eeq
\end{lem}
\begin{proof}
As in Proposition \ref{r1o2} we first  assume $B < Z^2$. The other case,
where the  terms have to be computed separately, works as in Lemma \ref{l11}.
Applying  Theorem \ref{IS1} with $\alpha =1$
and $d=1$ we get for arbitrary but fixed $\xpp$ (we set $\nu
=0$)
\beq
\left|\T (\zwei_1 \Theta_- (\wt H_{\xpp})) -
\int \dv \zwei_1 \Theta_- (\hp)\right| \leq
C\int_{{\rm supp} \zwei_1 (\xpp,z)} dz l(\x)^{-1}.
\eeq
This implies
\beq\label{2s3}
R^0(\zwei_1) \leq C\frac{B}{2\pi} \int d\x l(\x)^{-1}
\leq CB \int_{Z/B}^{cr_S} r^{
} dr \leq  C B[r_S]^{2}.
\eeq
\end{proof}

\begin{prop}\label{r2o2p}
\beq\label{r2o2}
\Rm_2(\Omega_2^1) \leq C B^{4/5} Z^{3/5}.
\eeq
\end{prop}
\begin{proof}
Note that by the Lemmata \ref{l11}, \ref{l12} and \ref{lem21} the estimate
(\ref{t11})
is proved and the assumption of Lemma \ref{l13} justified.
So by Lemma \ref{l13} and Lemma \ref{lem21} we arrive at
(\ref{r2o2}).
\end{proof}
\begin{prop}
\beq\label{r4o2}
\Rm_4 (\Omega_2^1), \Rm_3 (\Omega_2^1) \leq C B^{4/5} Z^{3/5}.
\eeq
\end{prop}
\begin{proof}
Let us start with
\beq
\Rm_4 (\Omega_2^1) = D(\zwei_1 (\rs - \bar \rho),\zwei_1 (\rs -
\bar
\rho)).
\eeq
By the HLS inequality we get
\beqa\nonumber
\Rm_4 (\Omega_2^1) & \leq& CB^2 \parallel \zwei_1 (\rs - \bar \rho)
\parallel_{6/5}^2\\
&\leq& C B^2 |\lambda_N - \nu|^2 \parallel \zwei_1
(\ps +\nu)^{-1/2}\parallel_{6/5}^2 \leq C B^{4/5} Z^{3/5}.
\eeqa
The term
\beq
\Rm_3(\Omega_2^1) = D(\zwei_1 (\rs -  \rho_\psi),\zwei_1 (\rs -
\rho_\psi))
\eeq
is a bit more delicate and we  refer to
\cite{Ivrii1} Proposition 4.2 and 4.3 for a proof of the estimate
(\ref{r4o2}).

{\bf Note:}
\, Proposition 4.3 in \cite{Ivrii1} is proved for region
$\chi_4=\{\x| |\x| \geq C_0 Z/B\}$ with possibly  a  very large
parameter $C_0$. This parameter $C_0$ is chosen in a way, such that
only the lowest Landau band contributes to Ivrii's calculations.
Since we only treat the lowest Landau band case
the assertion of Proposition 4.3 holds in our case  on the whole region
$\Omega_2^1$.

Furthermore, we remark that if (\ref{exx}) would be valid on
$\Omega_2^1$, we could immediately conclude by the HLS inequality
that $\Rm_4(\Omega_2^1) \leq CB^2[r_S]^3$. But since
the validity of (\ref{exx}) cannot be guaranteed on $\Omega_2^1$
we have to refer to Ivrii's method.
\end{proof}
Recall that we have made a partition of unity, $\sum_{i,j =1}^{2}
\psi^{(i)}_j (\x) =1$. So collecting all estimations of the
Sections 2.4 and 2.5 we have finished the proof of Theorem
\ref{qstf}.

\section{Semiclassical theories approximating $\Ec^{\rm Q}$}

As we have already argued throughout the introduction, the natural
semiclassical approximation of $\Ec^{\rm Q}$ is given by the
DSTF functional
\beq\label{dstff}
\E[\rho]=\sum_{m\in\N_0}\left(\kappa\int \rho_m(z)^3 - Z\int
V_m(z)\rho_m(z)dz\right)+\overline D(\rho,\rho).
\eeq
Here $\rho$ is a sequence of one-dimensional densities
$\rho= (\rho_m (z))_{m\in \N_0}$. In contrary to the usual STF
theory the integration over the variables orthogonal to the
magnetic field is replaced by an expansion in angular momentum
eigenfunctions in the lowest Landau band.
The potentials $V_m$ and $\ol D$ are defined in
(\ref{pot}).
The corresponding energy is given by
\beq\label{edstf}
E^{\rm DSTF}(N,Z,B)=\inf\left\{\E[\rho]\left| \ \rho \in \D \ \mbox{and} \ \
\sum_m\int \rho_m
\leq N\right.\right\},
\eeq
with
\beq
\D= \{ \rho | \ \sum_m \int\rho^{3}_m < \infty,\sum_m \int V_m \rho_m
< \infty, \widetilde D(\rho,\rho) <\infty\}.
\eeq
Another semiclassical approximation, where the variables,
as in the usual STF theory, are three dimensional densities,
is realized by the MSTF functional
\beq
\label{mstff} {\mathcal{E}} ^{\rm MSTF}[\rho ] =
\frac{4\pi^{4}}{3B^{2}} \int d\x \rho^{3}(\x) - \int d\x
\wt {|\x|^{-1}} \rho(\x) + \wt D(\rho ,\rho
),
\eeq
with respective energy
\beq\label{emstf}
E^{\rm MSTF}(N,Z,B)=\inf\left\{\Em [\rho]\left| \ \rho \in \wt \D \ \mbox{and} \ \
\int \rho
\leq N\right.\right\},
\eeq
where
\beq
\wt \D= \{ \rho | \ \rho \in \Ll^3(\R^3), \int \wt {|x|^{-1}} \rho
< \infty, \wt D(\rho,\rho) <\infty\}.
\eeq

First of all, we will show that these two functionals are
equivalent.

\begin{lem}\label{eqdm}
For all $\rho \in \wt \D$ let
\beq
\bar \rho(\x) = \frac B{2\pi} \sum_m \chi_m(\xpp) \rho_m(z),
\eeq
with $\rho_m (z) = \int \rho(\x)\chi_m(\xpp) d\xpp$,
and denote $\tilde \rho = (\rho_m)_m$.
Then one gets
\beq
\Em [\rho] \geq \Em [\bar \rho] = \E [\tilde \rho].
\eeq
\end{lem}
\begin{proof}
By the definition of the MSTF functional, it suffices to show
that $\int \rho^3 \geq \int \bar \rho^3$.

For this purpose we note that for every non-negative
function $f$, on a general measure space,
one derives from convexity that
\beq
\frac 1{\mu(\Omega)}\int f^3 d\mu \geq \left( \int \frac 1{\mu(\Omega)}
f d\mu \right )^3.
\eeq
Hence for every $m \in \N$ and $z \in \R$, we have
\beq
\frac 1{|\chi_m|} \int_{{\rm supp}\chi_m} \rho(\x)^3 \chi_m(\xpp) d\xpp \geq
\left( \frac 1{|\chi_m|} \int \rho(\x)\chi_m(\xpp) d\xpp \right)^3.
\eeq
Since $|\chi_m| = \frac {2\pi}B$, we arrive at
\beq
\frac{4\pi^4}{3B^2} \int \rho(\x)^3 d\x \geq \frac {\pi^2}3
\sum_m \int \rho_m(z)^3 dz = \frac{4\pi^4}{3B^2} \int \bar \rho(\x)^3
d\x.
\eeq
\end{proof}
\begin{prop}
For all $N,Z,B$
\beq
\Ems(N,Z,B) = \Ed(N,Z,B).
\eeq
\end{prop}
\begin{proof}
Lemma \ref{eqdm} immediately implies
\beq
\Ems(N,Z,B) = \inf \{ \Em [\rho]| \rho = \frac B{2\pi} \sum_m \rho_m(z)
\chi_m(\xpp),  \,\, (\rho_m)_m \in \D\}.
\eeq
\end{proof}
For simplicity we first concentrate on the DSTF functional and
then apply our results to the MSTF functional.
\begin{lem}\label{ubo}
$\E[\rho]$ is uniformly bounded from below on $\D$.There exists
even
a positive constant $\alpha$ and a $C$, such that
\beq
\E [\rho] \geq \alpha \left( \sum_m \int \rho_m^3 +
\ol D(\rho,\rho)\right) - C
\eeq
for all $\rho \in \D$.
\end{lem}
\begin{proof}
We set $\rho(\x) = \sum_m \rho_m(z) |\phi_m(\xpp)|^2$ for an arbitrary
$(\rho_m)_m \in \D$. We get from \cite{BBL} Lemma 2
that for every $\eps > 0$ there exists a $C_\eps$, such that
\beq
\int |\x|^{-1} \rho \leq \eps \parallel \rho \parallel_3 +
C_\eps D(\rho,\rho)^{1/2}.
\eeq
Hence, this implies
\beq\label{i31}
\sum_m \int V_m \rho_m \leq \eps \left( \int d\x \left( \sum_m
|\phi_m(\xpp)|^2\rho_m(z)\right)^3\right)^{1/3} + C_\eps \ol
D(\rho,\rho)^{1/2}.
\eeq
By convexity of $x^3$, for $x \geq 0$, and by the equation $\sum_m |\phi_m|^2 =
\frac B{2\pi}$, we get
\beq\label{i32}
\left( \frac {2\pi}B \sum_m |\phi_m|^2 \rho_m \right)^3
\leq \frac {2\pi}B \sum_m |\phi_m|^2 \rho_m^3.
\eeq
Using (\ref{i32}) and integrating over the $\xpp$-variable, the inequality (\ref{i31}) can
be written as
\beq
\sum_m \int V_m \rho_m \leq \eps (\frac B{2\pi})^{2/3} \left(
\sum_m \int
\rho_m^3\right)^{1/3} + C_\eps \ol
D(\rho,\rho)^{1/2}.
\eeq
Consequently the functional $\E[\rho]$ can be estimated from below
by $(\epsilon = \eps (\frac B{2\pi})^{2/3})$
\beqa\nonumber
\E[\rho] &\geq& \kappa \sum_m \int \rho_m^3 - \epsilon \left(\sum_m
\int\rho^3_m \right)^{1/3} + \ol D(\rho,\rho) - C_\eps \ol
D(\rho,\rho)^{1/2}\\
\nonumber &\geq& \inf_{X,Y \geq 0} \{ \kappa X^3 - \epsilon X + Y^2
- C_\eps Y\}\\
&\geq& \alpha (X^3 + Y^2) - C,
\eeqa
for $\alpha, C$ appropriately chosen,
where we used the notations $X = \left(\sum_m
\int\rho^3_m \right)^{1/3}$ and $Y = \ol
D(\rho,\rho)^{1/2}$
\end{proof}
\begin{lem}\label{exrho}
There exists a $\rho^{(\infty)}$, which minimizes $\E[\rho]$ uniquely
in $\D$, i.e. $\inf\{\E[\rho]| \rho \in \D\} = \E[\rho^{(\infty)}]$.
\end{lem}
\begin{proof}
Let $\rho^{(i)}$ be a minimizing sequence of $\E$.
Lemma \ref{ubo} yields that there exists a constant $C$, such that
\beq\label{exrho1}
\sum_m \kappa \int (\rho^{(i)}_m)^3 \leq C, \,\,
\ol D(\ri,\ri) \leq C, \,\, \sum_m \ri_m V_m \leq C
\eeq
for all $m \in \N_0$. By the Banach-Alaoglu theorem there exists a
subsequence, still denoted as $\ri$, and a $\rho^{(\infty)}$, with
$\ri_m \in \Ll^3(\R)\,\, \forall m \in \N_0$, such that
\beq\label{exrho2}
\ri_m \rightharpoonup \rho^{(\infty)}_m \,\, \mbox{weakly in} \,\,
\Ll^3(\R) \,\, \forall i
\in \N_0.
\eeq
Since $\Ll^p$-norms are weakly lower semicontinuous,
we derive for all $m$,
\beq
\liminf_{i\to\infty} \int (\ri_m)^3 \geq \int (\rho^{(\infty)}_m)^3,
\eeq
and using Fatou's Lemma we consequently arrive at
\beq
\liminf_{i\to\infty} \sum_m \int (\ri_m)^3 \geq
\sum_m \int (\rho^{(\infty)}_m)^3.
\eeq
Moreover since $V_m \in \Ll^{3/2}(\R)$ for all $m$, we conclude by
weak convergence
\beq
\lim_{i\to\infty} \int dz V_m (z)\ri_m (z) \to \int dz
V_m(z)\rho_m^{(\infty)}(z)
\eeq
for each $m$. By (\ref{exrho1}) and the dominated convergence theorem we have
\beq
\lim_{i\to\infty} \sum_m \int dz V_m(z)\ri_m(z) \to \sum _m \int dz
V_m(z)\rho_m^{(\infty)}(z).
\eeq
In order to show
\beq\label{exrho3}
\liminf_{i\to\infty} \ol D(\ri,\ri) \geq \ol
D(\rho^{(\infty)},\rho^{(\infty)}),
\eeq
we use the fact that for sequences of functions
$f=(f_m(z))_m$, $g=(g_m(z))_m$,
\beq
\langle f,g\rangle_D = \ol D(f,g)
\eeq
defines a real inner product and consequently a real
Hilbert-space ${ \mathcal{H}}_D$.
Since (\ref{exrho1}) yields $\parallel \ri \parallel_D =
\sqrt{\langle\ri,\ri\rangle_D} \leq C$ for all $i$,
we can extract another subsequence $\ri$, such that
\beq
\langle f, \ri\rangle \to \langle f, \rho^{(\infty)} \rangle
\,\, \mbox{for all} \,\, f \in { \mathcal{H}}_D.
\eeq
Hence, we conclude
\beqa\nonumber
&\ol D(\rho^{(\infty)}, \rho^{(\infty)}) = \lim_{i\to \infty} \langle\ri ,
\rho^{(\infty)} \rangle
\leq \langle \rho^{(\infty)},\rho^{(\infty)}\rangle^{1/2} \liminf_{i\to\infty}
\langle
\ri,\ri\rangle^{1/2}&\\& = \ol D(\rho^{(\infty)},\rho^{(\infty)})^{1/2}
\liminf_{i\to\infty} \ol D(\ri,\ri)^{1/2},&
\eeqa
and consequently get (\ref{exrho3}).
Altogether we have shown
\beq
\liminf_{i\to\infty}
\E[\ri] \geq \E[\rho^{(\infty)}].
\eeq
The uniqueness follows from the strict convexity of $\E$.
\end{proof}

\begin{thm}\label{thmds}
Denote $N_c = \sum_m \int \rho^{(\infty)}_m$. Then \\
(i) \,\, for each $N\leq N_c$  there exists a unique minimizer $\rho^{N}$ for
$\E$, under the restriction $\sum_m \int \rho_m \leq N$, i.e. $\Ed
(N,Z,B) = \E[\rho^{N}]$. Moreover, $\rho^{N}$ satisfies $\sum_m \int
\rho_m^{N} = N$.\\
(ii) \,\, $\Ed(N,Z,B)$, as a function of $N$, is strictly decreasing
and strictly convex up to $N_c$, and constant for $N > N_c$.
\end{thm}
\begin{proof}
Let $N \leq N_c$.
Then the  same proof as in Lemma \ref{exrho} shows that there
exists a $\rho^N \in \D$, with $\sum_m \int \rho^N_m \leq N$ and
\beq
\E[\rho^N] = \Ed(N,Z,B).
\eeq
Obviously $\Ed(N,Z,B)$, as a function of $N$, is non-increasing,
and the  convexity of $\E$ implies the convexity of $\Ed$. Hence,
by definition of $N_c$ and Lemma \ref{exrho} it is clear that
$\Ed$ is strictly decreasing up to $N_c$ and constant for $N >
N_c$. Furthermore we get that $\sum_m \int \rho^N_m = N$
for $N\leq N_c$. (Note that $\sum_m \int \rho^N_m < N$ would be a
contradiction to $N \leq N_c$.)
\end{proof}
\begin{prop}\label{propds}
Let $N \leq N_c$. Then for every minimizer $\rho^N$ there exists a
parameter $\mu(N)$, the chemical potential, such that $\rho^N$
obeys the coupled TF equations
\beq
3\kappa (\rho_m^{N}(z))^{2}= [ZV_m(z) - \sum_n \int
V_{m,n}(z-z')\rho_n^{N}(z') + \mu(N)]_+ \ \ \forall (m\in \N_0),
\eeq
and $\mu(N)$ fulfills the relation
\beq
\frac{\partial}{\partial N} \Ed(N,Z,B) = \mu(N).
\eeq
\end{prop}
\begin{proof}
The proof works analogously to \cite{LS} Theorem II.10,
if the variable perpendicular to the field is replaced by
the angular momentum quantum numbers.
\end{proof}
\begin{thm}
All statements of Theorem \ref{thmds} and Proposition \ref{propds}
are also valid for the MSTF theory, where the minimizing MSTF
densities $\rho^N(\x)$ and the minimizing DSTF densities
$(\rho^N_m(z))_m$ are related as
\beq
\rho^N(\x) = \frac B{2\pi} \sum_m \chi_m(\xpp)\rho^N_m(z).
\eeq
The corresponding TF equation reads
\beq
3\kappa (\rho^{N}(\x))^{2}= [Z\wt{|\x|^{-1}} - \sum_{n,m} \int d\x'
\chi_m(\xpp)
V_{m,n}(z-z')\chi_n(\xpp')\rho^{N}(\x') + \mu(N)]_+ .
\eeq
\end{thm}
\begin{proof}
The existence of a minimizing density $\rho^N(\x)$
we get from Theorem \ref{thmds} and Lemma \ref{eqdm}.
The uniqueness follows from the strict convexity of $\rho^3$ in $\rho$.
\end{proof}

Next we try to collect some information about the
\lq\lq critical\rq\rq\  particle number $N_c$, which measures the
maximal particle number that can be bound to the nucleus in the D(M)STF theory.

\begin{prop}\label{ncgz}
$N_c \geq Z$.
\end{prop}
\begin{proof}
By definition of $N_c$, we
have $\mu(N_c) =0$, so the TF equation reads ($\rho^{N_c} = \rho$)
\beq\label{007}
3\kappa \rho_m(z) = [\vph ]_+  \,\, \,\forall m \in \N_0,
\eeq
with
\beq
\vph = ZV_m(z)- \sum_n \int V_{m,n}(z-z')\rho_n(z').
\eeq
We assume $N_c < Z$.\\
 \cite{BRW,RW} tell us that the potentials $V_m(z)$ and
$V_{n,m}(z-z')$ behave like $1/|z|$ as $z \to \infty$. Hence, we get that
for each $m$
\beq
\lim_{z \to \infty} |z|[ZV_m(z)] = Z,
\eeq
as well as
\beqa\nonumber
&&\qquad\qquad\lim_{z \to \infty} |z|\left[ \sum_n \int V_{m,n}(z -
z')\rho_n(z')\right]\\ && =  \sum_n \int\lim_{z\to \infty} |z| V_{m,n}(z -
z')\rho_n(z') = \sum_n \int \rho_n = N_c.
\eeqa
Since we therefore get
\beq
\lim_{z \to \infty} |z|\vph = Z -N_c > 0,
\eeq
we can conclude that there exists an $\eps >0$ and a $\bar z >0$,
such that
\beq
\vph \geq \eps|z| \quad {\rm for} \quad z \geq \bar z,
\eeq
which by (\ref{007}) is a contradiction to $\rho_m \in \Ll^1(\R)$.
\end{proof}

In the usual STF theory the inequality $N_c \leq Z$  is a consequence of
Newton's potential-theory. Since we miss this powerful tool in our
DSTF theory we cannot expect to get an analogue estimate.
But if we use similar methods to those applied in \cite{BRW,Sei,HS}
we at least get the following $B$-independent upper bound for
$N_c$.

\begin{prop}
$N_c \leq 4Z$.
\end{prop}
\begin{proof}
If we multiply (\ref{007}) with
$\rho_m/V_m$ and integrate over $z$, we get
\beq
3\kappa\int dz \frac{\rho_m(z)^3}{V_m(z)} = Z\int dz \rho_m(z) - \sum_n
\int dz dz'\frac 1{V_m(z)} \rho_n(z') V_{n,m}(z-z')\rho_m(z).
\eeq
Note that by multiplication with $\rho_m$ the $[]_+$-bracket
can be dropped, since $\rho_m =0$ where $\vph \leq 0$.
Clearly $\int \rho_m^3/V_m \geq 0$, so after summing over
$m$ we arrive at
\beq\label{008}
ZN_c \geq
\sum_{n,m}
\int dz dz'\frac 1{V_m(z)} \rho_n(z') V_{n,m}(z-z')\rho_m(z).
\eeq
Moreover, \cite{HS} Lemma 4.1 tells us
\beq\label{assert}
\left(\frac 1{V_m(z)}+\frac 1{V_n(z)}+\frac 1{V_m(z')}+\frac 1{V_n(z')}\right)
V_{m,n}(z-z')\geq 1,
\eeq
which we use, together with symmetry, in order to estimate the right side of
(\ref{008}):
\beqa\nonumber
&&\sum_{m,n}\int \frac 1{V_m(z)}\rho_m(z) V_{m,n}(z-z')
\rho_n(z') d zd z' \\ \nonumber &&=\frac 14 \sum_{m,n}\int
\left(\frac 1{V_m(z)}+\frac 1{V_n(z)}+\frac 1{V_m(z')} +\frac
1{V_n(z')}\right)\\ \label{nci} &&\qquad \qquad \times
\rho_m(z)V_{m,n}(z-z')\rho_n(z') d zd z' \geq \frac 14
N_c^2.
\eeqa
Inserting into (\ref{008}) finally leads to
\beq
N_c \leq 4Z.
\eeq
\end{proof}

\begin{rem}[The difference between $E^{\rm DSTF}$ and $E^{\rm STF}$]

Obviously, the magnitude of difference between the D(M)STF and the STF energy
is given by
\beq\label{rem11}
B\left[\int |\psm(\x)|^{3/2} - \int |\ps(\x)|^{3/2}\right].
\eeq
Due to the singularity of the STF potential, (\ref{rem11}) has to be
split into
\beqa\nonumber &&
B \int_{|\x| \leq B^{-1/2}}\left[ |\psm(\x)|^{3/2} - \int |\ps(\x)|^{3/2}\right]
\\ \label{rem2}
&& \qquad \qquad+ B\int_{|\x| \geq B^{-1/2}}\left[ |\psm(\x)|^{3/2} -
\int |\ps(\x)|^{3/2}\right].
\eeqa
The magnitude of the first term is proportional to
\beq
Z^{3/2}B\int_{|\x| \leq B^{-1/2}} |\x|^{-3/2} = O(Z^{3/2}B^{1/4}).
\eeq
The second term of (\ref{rem2}) could be estimated by

\beq\label{rem3}
 B\int_{|\x| \geq B^{-1/2}}|\ps(\x)|^{1/2} |\partial_\theta \ps(\x) |B^{-1/2}
\leq Z^{3/2}B^{1/2}[r_S]^{3/2},
\eeq
with $\theta = |\xpp|$. So we see that  the main contribution to (\ref{rem11})
stems from the $B^{-1/2}$-vicinity of the nucleus, i.e.
\beq
E^{\rm DSTF} - E^{\rm STF} = O(Z^{3/2}B^{1/4}).
\eeq
\end{rem}

\subsection{Some notes about the one-dimensional case}

If we reduce the DSTF functional to the angular momentum channel
with $m=0$, one gets the functional
\beq\label{1dstf}
\Edo [\rho] = \kappa\int dz \rho(z)^3 - Z\int dz V_0 (z) \rho(z)
+\frac 12 \int dz dz' V_{0,0} (z-z')\rho(z) \rho(z'),
\eeq
which can be treated analogously to the three dimensional case
and Theorem \ref{thmds} and Proposition \ref{ncgz}
are also valid. Concerning the upper bound of $N_c$ it is not
necessary to symmetrize over $n$ and $m$, and in this case
(\ref{nci}) reads
\beqa\nonumber
&&\int \frac 1{V_0(z)}\rho(z) V_{0,0}(z-z')
\rho(z') d zd z' \\ \nonumber &&=\frac 12 \int
\left(\frac 1{V_0(z)}+ \frac 1{V_0(z')}\right)
\rho(z)V_{0,0}(z-z')\rho(z') d zd z' \geq \frac 12
N_c^2.
\eeqa
Consequently one gets $N_c \leq 2Z$ for the maximum particle
number that can be bound to the nucleus in the one-dimensional theory.

Moreover, let us regard the absolute minimum $\bar E^{1\rm DSTF}(Z,B)$
of the functional  $\Edo [\rho] = \Ema_{Z,B}[\rho]$.
If we use the scaling relations
\beq
V_0 (z) = B^{1/2}V_0^1(B^{1/2}z), \quad
V_{0,0} (z) = B^{1/2}V_{0,0}^1 (B^{1/2}z)
\eeq
and define
\beq\label{scalr}
\bar \rho(z) = B^{1/4}Z^{1/2} \rho(B^{1/2} z)
\eeq
we get
\beq\label{lam}
\Ema_{Z,B}[\bar \rho] = B^{1/4}Z^{3/2} \Ema_{1,1}^{\lambda}[\rho],
\eeq
with
\beq\label{11}
\Ema_{1,1}^{\lambda} [\rho] = \kappa\int dz \rho(z)^3 - \int dz V^1_0 (z)
\rho(z)
+\frac 1\lambda \int dz dz' V_{0,0}^1(z-z') \rho(z) \rho(z'),
\eeq
and $\lambda = 2B^{1/4}Z^{1/2}$.

Let $E^{1D}_w (Z,B)$ be the minimum of the functional
\beq\label{1Df}
\Ema_w^{1D} [\rho] = \kappa\int dz \rho(z)^3 - Z\int dz V_0 (z)
\rho(z),
\eeq
where the repulsive energy term is omitted.
Using the above scaling (\ref{scalr}) one immediately gets $E_w^{1D}(Z,B)
= Z^{3/2}B^{1/4}
E^{1D}_w(1,1)$. So we can formulate the following theorem:
\begin{thm}\label{oneth}
If $Z,B \geq1$ are fixed, then
\beq
E^{1D}_w (Z,B)  \leq  \bar E^{1\rm DSTF}(Z,B) \leq  E^{1D}_w (Z,B)
+ Z(1 + 2\ln[BZ^2]^2).
\eeq
\end{thm}
\begin{proof}
The lower bound is obvious.

For the upper bound we use the relation (\ref{lam}) and
take the TF-solution of $\Ema_{1,1}^{\infty}$,
i.e.
\beq
\rho(z) = \frac 1\pi \sqrt{V_0^1(z)}.
\eeq
This density is neither in $\Ll^1$ nor in $\Ll^2$, so we define a
cut-off density $\rho_R(z) = \pi \sqrt{V_0^1(z)} \Theta (R-|z|)$
and use this as comparison density in (\ref{lam}), which leads to
\beq\label{split}
\Ema_{1,1}^{\lambda} [\rho_R] = E^{1D}_w (1,1) + \int_R^\infty (V_0^1(z))^{3/2}
+\frac 1\lambda \int \rho_R(z) V_{0,0}^1(z - z') \rho_R(z') dzdz'.
\eeq
Since $V_{0,0}^1(z) \leq \min\{\frac 1{|x|}, \sqrt{\pi/4}\}$ we get by
Young's inequality
\beq\label{opt}
\int \rho_R(z) V_{0,0}^1(z - z') \rho_R(z') dzdz' \leq
\left[\left(\int \rho_R\right)^2 \frac 1\beta + 2\ln(\beta)\int \rho_R^2\right]
\quad
\forall \beta \geq 1.
\eeq
After estimating $\int \rho_R$ and $\int \rho^2_R$ we see that
the minimum of (\ref{opt}) as a function of $\beta$ is achieved for
$\beta= R/\ln(R)$, which implies
\beq
\int \rho_R(z) V_{0,0}^1(z - z') \rho_R(z') dzdz' \leq
[\ln(R) + \ln(R)^2].
\eeq
Next, optimizing the last two term on the right side
of (\ref{split}) with respect to $R$ and multiplying with $B^{1/4}Z^{3/2}$
yields the statement of the theorem.
\end{proof}
By aid of this theorem we can also prove that $\bar E^{1\rm DSTF}(Z,B)$
is the semiclassical approximation of $\T[-\partial_z^2 - ZV_0(z)]_-$,
the sum of all negative eigenvalues of $-\partial_z^2 -ZV_0(z)$.
\begin{cl}
Let $B,Z \geq 1$ and $B \leq Z^2$. Then there exists a constant $C$, such that
\beq
|\T[-\partial_z^2 - ZV_0(z)]_- - \bar E^{1\rm DSTF}(Z,B)| \leq C\max\{Z\ln[BZ^2],
B^{3/4} Z^{1/2}\}.
\eeq
\end{cl}
\begin{proof}
This is an immediate consequence of Theorem \ref{oneth} and \cite{H1}
Theorem 3.19, which says that
\beq
|\T[-\partial_z^2 - ZV_0(z)]_- - E_w^{1D}(Z,B)| \leq C
B^{3/4} Z^{1/2}.
\eeq
\end{proof}

We learn from Theorem 3.10 that in a model of a one dimensional semiclassical
atom, where the electrons are forced to stay in the angular momentum channel
$m=0$, the repulsive interaction energy does not contribute to the leading order
of the energy $\bar E^{1\rm DSTF}(Z,B)$ for large $Z$ and $B \geq 1$.

An analogue effect one obtains for the quantum mechanical interaction energy of
$N$ particles reduced to the angular momentum $m=0$,
i.e.
\beq\label{mo}
\Psi = \phi_0 \otimes...\otimes \phi_0 \psi(z_1,...z_N).
\eeq
For $\psi$ a Slater-determinant or at least for $\psi$ close to the ground state of
the corresponding $N$-particle Hamiltonian $H_0$, which is the projection onto
the angular momentum eigenspace with angular momentum $m=0$,
the interaction energy
can be bounded from above by (for a precise lower bound see \cite{HS2})
\beq
\frac 12 \int_{\R^2} \rho_\psi (z)\rho_\psi(z') V_{0,0}(z-z')dzdz',
\eeq
which can be estimated by an  analogue method to (\ref{opt}).
This leads to
\beq\label{370}
\frac 12 \int_{\R^2} \rho_\psi (z)\rho_\psi(z') V_{0,0}(z-z')dzdz'
\leq C E_0^{1/2} N^{1/2} \left[1 + \ln (BN^3/E_0)\right],
\eeq
where we have used that $\langle \Psi, H_N\Psi\rangle \leq 0$
and $E_0$ is the corresponding ground state energy
(of wave functions of the form (\ref{mo})), which is of the same order
as $\bar E^{1\rm DSTF}$ as long as $B \leq Z^2$.
Relation (\ref{370}) yields that the quantum mechanical interaction energy
in one dimension is $\ll E_0$ as long as $E_0 \gg N$.\\ \\
\noindent
{\bf Acknowledgement.} The author thanks Jakob Yngvason for proofreading and his friend Robert W. Seiringer
for many helpful discussions and comments.

\end{document}